# Efficient fluorescence quenching in electrochemically exfoliated graphene decorated with gold nanoparticles

M. Hurtado-Morales[1], M. Ortiz[1], C. Acuña[1], H.C. Nerl[2], V. Nicolosi[3], Y. Hernández[1*]

[1]Nanomaterials Laboratory, Physics Department, Universidad de los Andes, Bogotá – Colombia.

[2] CRANN & AMBER and School of Physics, Trinity College Dublin, Dublin 2, Ireland

[3] CRANN & AMBER, School of Physics & School of Chemistry, Trinity College Dublin, Dublin 2, Ireland

## Abstract

High surface area graphene sheets were obtained by electrochemical exfoliation of graphite in an acid media under constant potential conditions. Filtration and centrifugation processes played an important role in order to obtain stable dispersions in water. Scanning electron microscopy (SEM) and transmission electron microscopy (TEM) imaging revealed highly exfoliated crystalline samples of ~5μm. Raman, FT-IR and XPS spectroscopy further confirmed the high quality of the exfoliated material. The electrochemically exfoliated graphene (EEG) was decorated with gold nanoparticles (AuNP) using sodium cholate (SC) as a buffer layer. This approach allowed for a non-covalent functionalization without altering the desirable electronic properties of the EEG. The AuNP-EEG samples were characterized with various techniques including absorbance and fluorescence spectroscopy. These samples displayed a fluorescence signal using an excitation wavelength of 290nm. The calculated quantum yield (Φ) for these samples was 40.04%, a high efficiency compared to previous studies using solution processable graphene.

* yr.hernandez@uniandes.edu.co

## 1. Introduction

Liquid phase exfoliation of graphite to produce graphene [1-3] has shown to be an efficient processing method when designing new applications [4]. In recent years, electrochemical exfoliation has proved to be an adequate method to produce a high quality graphene solution, useful for applications such as lithium storage [5], supercapacitors [6], photodetectors [7] and transparent electrodes [8]. The produced samples are a few layers thick (~3 layers), with large crystal sizes (a few microns) and with high electrical conductivity [8].

Fluorescent biosensors could take advantage of the high electrical/thermal conductivity of graphene as well as its chemical and mechanical stability [9, 10]. One of the greatest challenges in order to design graphene-based biosensors is the ability to guarantee a stable binding between the surface and the fluorescent molecule [11]. Previous reports on graphene-based biosensors have used strong oxidation methods in order to locate functional groups on the surface that allow decoration with a variety of nanoparticles [12]. Other studies that measured quantum yield of decorated graphene (not graphene oxide) reported values of 10.54%, 0.93% and 42.66% for three dyed molecules respectively: Rhodamine B, eosin and methylene blue deposited on non-covalently functionalized graphene sheets [13]. Further reports have optically characterized dye-tagged DNA molecules on graphene for biomolecular detection [10]. Additionally, the interaction of graphene with metal surfaces has also proved to enhance the sensitivity of graphene biosensors due to surface plasmon resonance effects [14, 15].

In this work, a green electrochemical potentiostatic method in acid media is used in order to exfoliate graphite to obtain large area graphene sheets. The produced graphene is readily dispersible in water, which is advantageous for device processing. A non-covalent functionalization process of EEG with AuNP was designed using SC as a buffer layer. The optical properties of the produced composites were studied using absorbance and fluorescence spectroscopy. These studies allowed the calculation of the quantum yield, which provides the fluorescence quenching efficiency.

## 2. Experimental

For the experiment, graphite foil (99.8%) was purchased from Alfa Aesar; fume sulfuric acid, isopropanol and SC, from Sigma-Aldrich; aqueous dispersion of AuNP, with a diameter of 14nm, from US Nanomaterials Research. For the electrochemical exfoliation process, a lab-built potentiostat was designed to allow output voltages of up to $\pm$10V. For our purpose, a Pt wire was used as counter electrode and a 1cm x 3cm piece of carbon foil played the role of working electrode (supporting information Fig. S1).

Sulfuric acid at 0.1M was used to carry out the exfoliation process as previously reported by Muellen and coworkers[8]. An effective area of 2cm$^2$ of the carbon foil was in direct contact with the acid media. The electrochemical reaction was carried out at +8V. During the reaction, graphite flakes peeled off from the working electrode until the area in contact with the acid media was consumed (~25 min).

The carbon material left in the acid media after the electrochemical reaction was rinsed and redispersed in 40ml of mili-Q water, using an ultrasonic bath (Branson® 5800) for 90 minutes. Then, a centrifugation process (at 3500rpm for 30 minutes) was used to remove the non-exfoliated graphite material. Afterwards, the supernatant was ultrasonicated during 90 minutes to assure a uniform dispersion of the produced graphene. The concentration of the dispersion was quantified via optical absorption spectroscopy, performed using an Analytik Jena Specord®50 UV-vis-NIR spectrometer.

Surfactant molecules have been used extensively to de-bundle carbon nanotubes [16] and to exfoliate graphite to produce graphene [17]. In this case, SC was used not only to stabilize graphene in water, but also to non-covalently functionalize the surface of the EEG. A 2% wt of SC was added to the EEG-water dispersion followed by 120 minutes of ultrasonication. Subsequently, AuNP suspensions of 1.70mM, 1.02mM, 0.72mM and 0.56mM concentration were added to the EEG/SC heterostructure.

## 3. Results

The structural characterization of the EEG was performed using microscopic and spectroscopic techniques. Figure 1a shows a representative SEM image of EEG sheets deposited on a $SiO_2$/Si substrate by spray coating. Even though the EEG can be produced from various graphite sources [18], we found that the quality of the starting material has a direct effect on the size and crystallinity of the exfoliated graphene. Figure 1b corresponds to a representative TEM image of EEG sheets on a holey carbon TEM grid. In both images, it is clear that the EEG is composed by a few layers of graphene with incommensurate stacking, as confirmed by Selected Area Electron Diffraction (SAED). These results are further corroborated by AFM imaging (supporting information Fig. S3).

Raman spectroscopy of the EEG revealed a sharp D peak, at 1349.6$cm^{-1}$, which is characteristic of edges or vacancies on the graphene surface [8]. The intensity of the G peak, at 1583.9 $cm^{-1}$, remains high when compared to the 2D peak, confirming that the flakes are a few layers thick. The second order Raman peaks at around 2441$cm^{-1}$; 2936$cm^{-1}$ and 3239$cm^{-1}$ are specific bands that appear for disordered carbon and nanostructured carbons [19]. The peak at around 2707$cm^{-1}$ (2D peak) represents the second harmonic of the D peak, which is symmetric and very intense in comparison to the first-order G peak in single layer graphene or thinly stacked graphitic structures [20]. On the other hand, the 2D peak in pure crystalline graphite is typically less intense and splits in two separate peaks. The ratio between the intensity of D and G peaks is often used to quantify the amount of defects in graphitic materials and can be related to the average distance between defects using the Tuinstra-Koening ratio [21]. This ratio was found to be $I_D/I_G$=0.78 in our samples, leading to an average inter-defect distance of ~27nm.

The appearance of a shoulder at the 2D peak indicates a small level of AB stacking within the exfoliated material [22]. The position of the D peak depends on the energy of the laser, but contrary to the 2D band, the D peak does not depend on the number of layers; it depends on the amount of disorder. The lateral size of the graphene flake is about 5μm; given that the samples were deposited on $SiO_2$/Si substrates, we attribute the presence of the D peak to wrinkles, edges or vacancies on the basal plane.

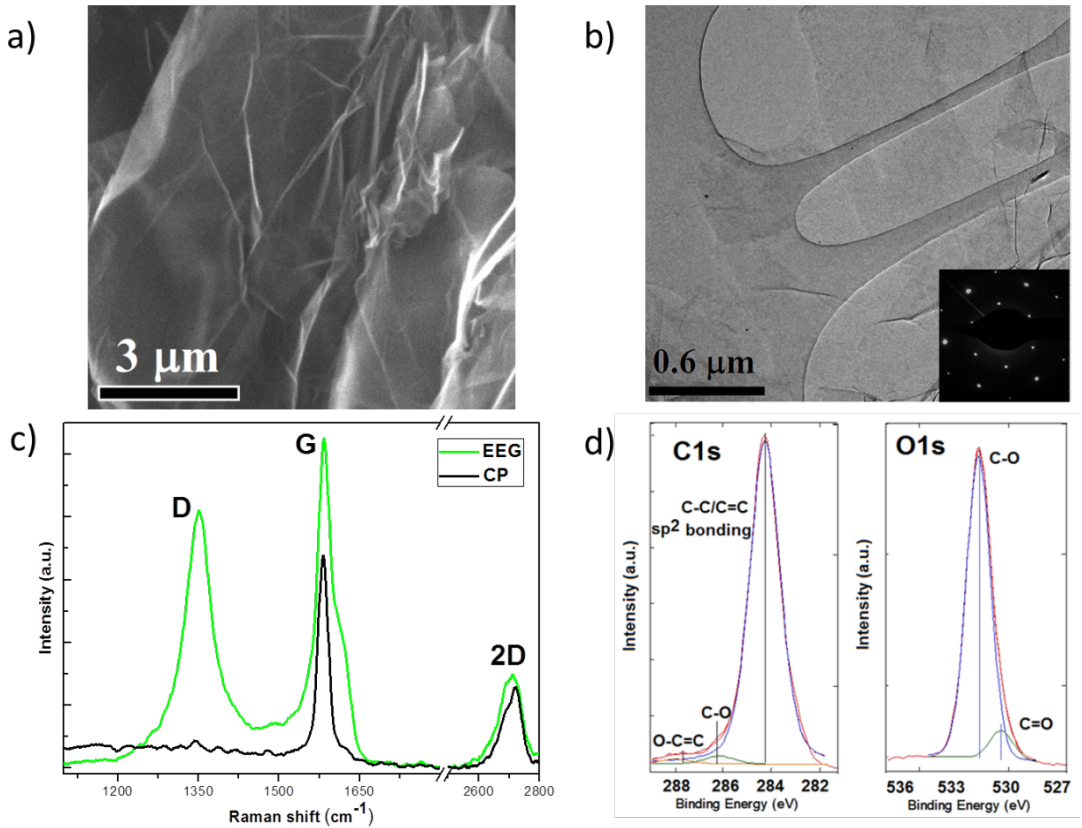

Fig. 1. a) SEM image of EEG sheets on a $SiO_2$/Si substrate. b) TEM image of EEG and SAED analysis. c) Carbon foil (CF) and EEG Raman spectra (Laser 532nm). d) XPS spectra of the EEG C1s and O1s spin-orbital couplings.

High resolution XPS measurements were carried out in a HV chamber, equipped with a focused monochromatized X-ray source (AlKα=1486.6eV), to probe the surface composition and chemical state of an EEG sample, which was deposited on a $SiO_2$/Si substrate by spray coating. Figure 1d shows that less than 2.3% of oxygen is present in the exfoliated graphene after the electrochemical process. The resolved fitted XPS spectra of the C 1s signal (284.0eV) shows the spin-orbital coupling from C-C, being this signal close to 100%. Furthermore, a residual contribution of C-OH (286.4 eV) and C(O)-O (289.1 eV) functional groups was observed in the graphene structure [23-26]. These signals can be attributed to residual acid at the EEG edges and gas absorption on the sample. The absence of a strong C=O peak, at 1717cm$^{-1}$, in the XPS is consistent with our FTIR analysis (Supporting information Fig. S4).

**Non-Covalent Graphene decoration**

In order to study the success of the decoration process, the morphology of the samples was studied using electron microscopy. The TEM image in Figure 2a is representative of an AuNP-EEG flake, while the TEM image in Figure 2b shows an individual AuNP at higher magnification. Meanwhile, Figure 2c is a representative SEM image of the decorated samples. Lastly, the Scanning TEM (STEM) image in Figure 2d clearly shows the distribution of the AuNp on an EEG flake. Since the Au atoms scatter the beam electrons more strongly than carbon atoms do, they appear much brighter in the STEM image. EDX analysis corroborates the presence of AuNP on the EEG surface (Supporting information Fig. S5). It is also important to note that the AuNP observed in the TEM are all single particles, not clusters. This type of non-covalent decoration of graphene is mediated by the electrostatic interaction with the SC molecule. Preliminary results proved that simple addition of AuNP to EEG dispersions does not produce AuNP decorated graphene.

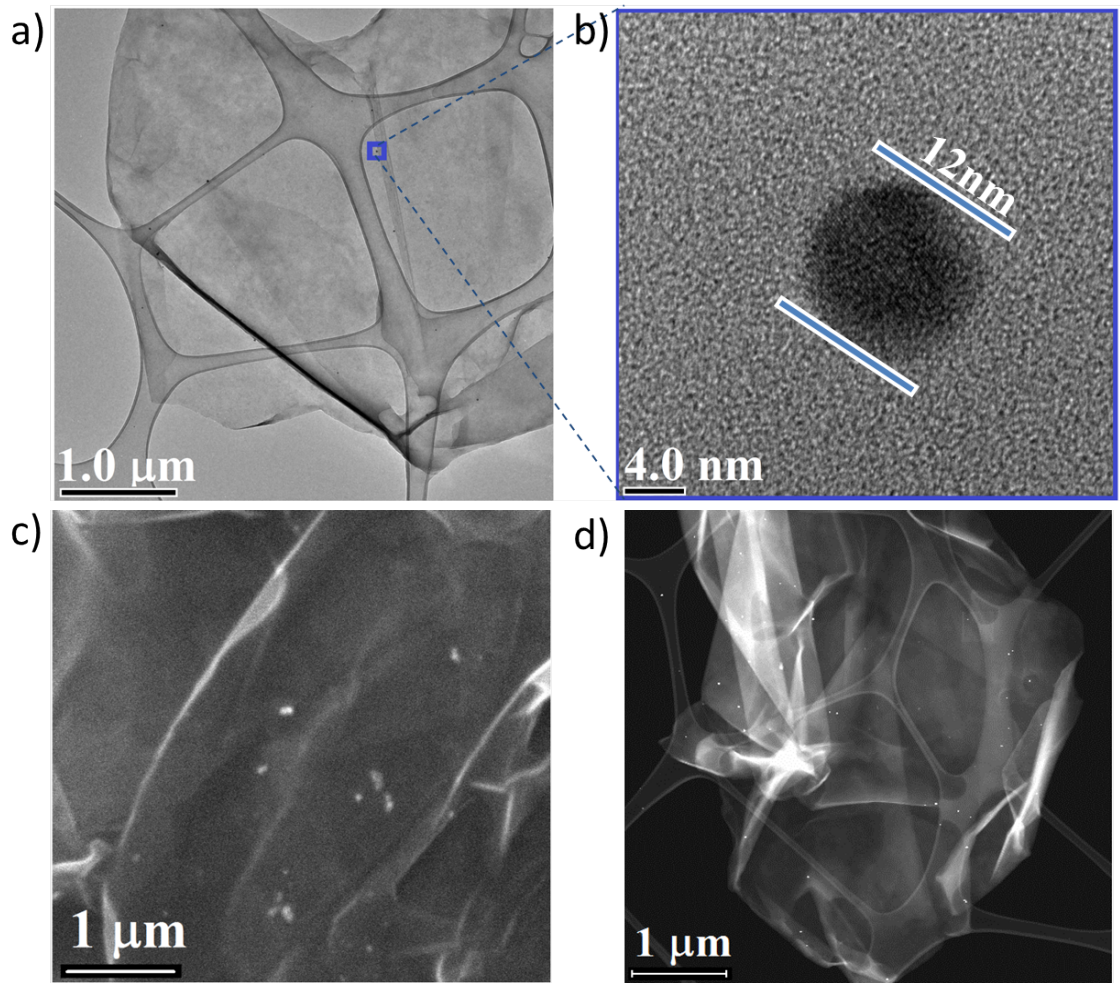

Fig. 2. a) TEM image of a AuNP-EEG heterostructure. b) AuNP on the surface of EEG. c) SEM image of a AuNP-EEG composite. d) Scanning TEM image of a AuNP-EEG heterostructure.

Raman spectroscopy was performed on the produced composites AuNP-SC-EEG using two different laser lines (Fig. 3a). In the case of the 638nm laser, the D and G peaks were observed with an intensity ratio of $I_D/I_G$=1.23, a similar value for the one measured using the 532nm laser $I_D/I_G$=1.26. It is noteworthy that $I_D/I_G$ is higher in both cases when compared to the non-functionalized graphene, leading to an inter-defect distance of ~2nm.

Furthermore, peak shifts were observed in the AuNP-SC-EEG samples; when using a 532nm laser, the D peak was located at 1342.2cm$^{-1}$, while the G peak was located at 1586.5cm$^{-1}$. In this case, the D and G peaks are shifted 7.39cm$^{-1}$ to the left and 2.58cm$^{-1}$ to the right, respectively. In the case of a 638 nm laser, the D and G peaks are located at 1333.8 cm$^{-1}$ and 1592.9 cm$^{-1}$, respectively.

Because of the presence of SC molecules and AuNP, the G peak is reduced in intensity when compared with the D peak, causing at the same time bathochromic shifts of 3.7cm$^{-1}$ and 4.7cm$^{-1}$ for the D and G peaks, respectively. On the other hand, measurements carried out with a 532nm laser showed two different shifts; the first one is a bathochromic shift, of 7.4cm$^{-1}$, on the D peak, with respect to the same peak from a non-functionalized sample. The second one, a hypsochromic shift, of 2.6cm$^{-1}$ on the G peak, when compared to the same band from the non-functionalized sample. The red-shifted peaks can be associated to the n-doping effect induced by gold nanoparticles [27-29].

Additionally, AuNP-SC-EEG composites were characterized with UV-vis optical absorption. Figure 3b illustrates the observed and expected optical absorption of the composites prepared with different AuNP concentrations. It is noteworthy that there are no visible peak shifts in the decorated samples, when compared to the as-received material that could be associated to structural changes in the nanoparticles. The UV-vis optical absorption of the composites (solid lines) was under the expected values (dashed lines), except for the highest concentration of AuNP. This anomalous effect could be attributed to the reabsorption of light by the AuNP, within the solution that interferes with the signal of the fluorescence of the aqueous dispersion.

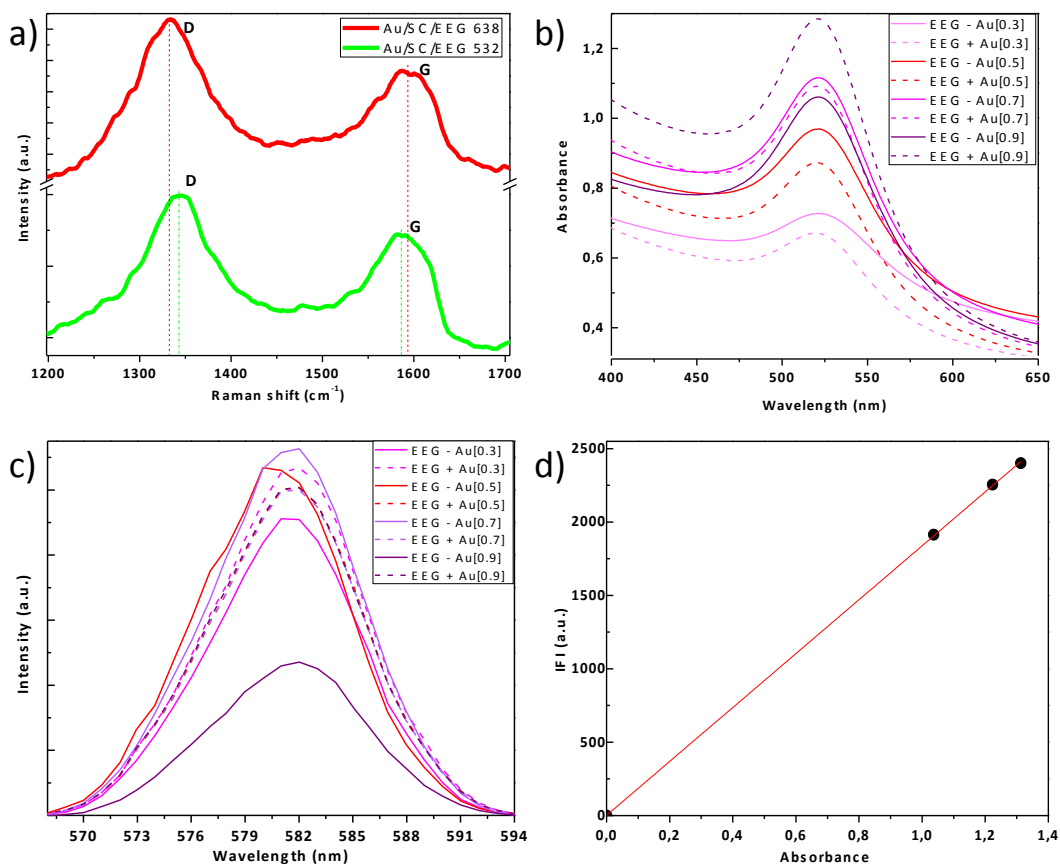

Fig.3. a) Raman shifts AuNp-SC-EEG. b) UV-vis optical absorption of AuNp-SC-EEG composites. Color gradient points to gradient concentrations; the darker shades designate higher concentrations, while the lighter shades designate lower concentrations. Solid and dashed lines describe observed and expected emission for each respective concentration. c) Fluorescence emission of AuNp-SC-EEG composites. d) High correlation between the absorbance and the integrated fluorescence intensity (IFI) of every decoration carried out for the Φ determination.

The Quantum Yield (Φ) is obtained by extracting the slope from the linear fit of the absorbance vs. the integrated fluorescence intensity (IFI) graph of the composite to be tested [30]. The quantum yield of the composite is calculated using a reference fluorophore of known Φ. In this study, the reference sample was Rhodamine B with Φ = 71% [31]. Figure 3d shows the lineal relation between both optical parameters from which the calculated quantum yield is 40.04%. One of the highest values reported so far for solution exfoliated graphene [13]. The EEG process preserves the high electrical and thermal conductivity of graphene, which combined with the large area of the produced crystals is favorable for fluorescence quenching.

## 4. Conclusions

High surface area graphene sheets were obtained by electrochemical exfoliation of graphite in acid media, under constant potential conditions. SEM and TEM images revealed highly exfoliated samples with crystal sizes of ~5μm. Meanwhile, FTIR and XPS spectroscopy confirmed the high quality of the exfoliated material. The aim of the study was to decorate the EEG with AuNP, using SC as a buffer layer, which allows a non-covalent functionalization that preserves its electronic structure. Fluorescence and Raman spectroscopy combined with HRTEM showed that the decoration was successful, leading to a calculated quantum yield of 40.04%. One of the highest compared to previous reports of solution processable graphene.

## References

1.   Hernandez, Y., et al., *High-yield production of graphene by liquid-phase exfoliation of graphite.* Nature Nanotechnology, 2008. **3**(9): p. 563-568.

2.   Lotya, M., et al., *Liquid Phase Production of Graphene by Exfoliation of Graphite in Surfactant/Water Solutions.* Journal of the American Chemical Society, 2009. **131**(10): p. 3611-3620.

3.   Stankovich, S., et al., *Graphene-based composite materials.* Nature, 2006. **442**(7100): p. 282-286.

4.   Coleman, J.N., *Liquid Exfoliation of Defect-Free Graphene.* Accounts of Chemical Research, 2013. **46**(1): p. 14-22.

5.   Lung-Hao Hu, B., et al., *Graphene-modified LiFePO4 cathode for lithium ion battery beyond theoretical capacity.* Nat Commun, 2013. **4**: p. 1687.

6.   Wu, Z.-S., et al., *Ultrathin Printable Graphene Supercapacitors with AC Line-Filtering Performance.* Advanced Materials, 2015. **27**(24): p. 3669-3675.

7.   Liu, Z., et al., *Transparent Conductive Electrodes from Graphene/PEDOT:PSS Hybrid Inks for Ultrathin Organic Photodetectors.* Advanced Materials, 2015. **27**(4): p. 669-675.

8. Parvez, K., et al., *Electrochemically exfoliated graphene as solution-processable, highly conductive electrodes for organic electronics.* ACS Nano, 2013. **7**(4): p. 3598-3606.

9. Song, Y., et al., *Recent advances in electrochemical biosensors based on graphene two-dimensional nanomaterials.* Biosensors and Bioelectronics, 2016. **76**: p. 195-212.

10. Hong, B.J., et al., *Tunable Biomolecular Interaction and Fluorescence Quenching Ability of Graphene Oxide: Application to "Turn-on" DNA Sensing in Biological Media.* Small, 2012. **8**(16): p. 2469-2476.

11. Tang, L., Y. Wang, and J. Li, *The graphene/nucleic acid nanobiointerface.* Chemical Society Reviews, 2015. **44**(19): p. 6954-6980.

12. Wang, Z., et al., *Comparative studies on single-layer reduced graphene oxide films obtained by electrochemical reduction and hydrazine vapor reduction.* Nanoscale Research Letters, 2012. **7**(1): p. 161-161.

13. Liu, Y., C.-y. Liu, and Y. Liu, *Investigation on fluorescence quenching of dyes by graphite oxide and graphene.* Applied Surface Science, 2011. **257**(13): p. 5513-5518.

14. Wu, L., et al., *Highly sensitive graphene biosensors based on surface plasmon resonance.* Optics Express, 2010. **18**(14): p. 14395-14400.

15. Pau, J.L., et al., *Optical biosensing platforms based on Ga–graphene plasmonic structures on Cu, quartz and SiO2/Si substrates.* physica status solidi (b), 2016: p. n/a-n/a.

16. Moore, V.C., et al., *Individually suspended single-walled carbon nanotubes in various surfactants.* Nano Letters, 2003. **3**(10): p. 1379-1382.

17. Coleman, J.N., *Liquid-Phase Exfoliation of Nanotubes and Graphene.* Advanced Functional Materials, 2009. **19**(23): p. 3680-3695.

18. Singh, V.V., et al., *Greener Electrochemical Synthesis of High Quality Graphene Nanosheets Directly from Pencil and its SPR Sensing Application.* Advanced Functional Materials, 2012. **22**(11): p. 2352-2362.

19. Hiura, H., et al., *Raman studies of carbon nanotubes.* Chemical Physics Letters, 1993. **202**(6): p. 509-512.

20. Pimenta, M.A., et al., *Studying disorder in graphite-based systems by Raman spectroscopy.* Physical Chemistry Chemical Physics, 2007. **9**(11): p. 1276-1290.

21. Ferrari, A.C. and D.M. Basko, *Raman spectroscopy as a versatile tool for studying the properties of graphene.* Nat Nano, 2013. **8**(4): p. 235-246.

22. Ferrari, A.C., et al., *Raman Spectrum of Graphene and Graphene Layers.* Physical Review Letters, 2006. **97**(18): p. 187401.

23. Li, J., et al., *Experimental investigation of the important influence of pretreatment process of thermally exfoliated graphene on their microstructure and supercapacitor performance.* Electrochimica Acta, 2015. **180**: p. 187-195.

24. Zhai, P., et al., *Water-soluble Microwave-exfoliated Graphene Nanosheet/Platinum Nanoparticle Composite and Its Application in Dye-Sensitized Solar Cells.* Electrochimica Acta, 2014. **132**: p. 186-192.

25. Antiohos, D., et al., *Manganosite–microwave exfoliated graphene oxide composites for asymmetric supercapacitor device applications.* Electrochimica Acta, 2013. **101**: p. 99-108.

26. Song, Y., J.-L. Xu, and X.-X. Liu, *Electrochemical anchoring of dual doping polypyrrole on graphene sheets partially exfoliated from graphite foil for high-performance supercapacitor electrode.* Journal of Power Sources, 2014. **249**: p. 48-58.

27. Liu, M., et al., *Interface Engineering Catalytic Graphene for Smart Colorimetric Biosensing.* ACS Nano, 2012. **6**(4): p. 3142-3151.

28. Shaban, M., A.G.A. Hady, and M. Serry, *A New Sensor for Heavy Metals Detection in Aqueous Media.* Sensors Journal, IEEE, 2014. **14**(2): p. 436-441.

29. Zhao, L., et al., *In situ regulation nanoarchitecture of Au nanoparticles/reduced graphene oxide colloid for sensitive and selective SERS detection of lead ions.* Journal of Colloid and Interface Science, 2016. **465**: p. 279-285.

30. Yvon, H.J., *A guide to recording Fluorescence Quantum Yields.* HORIBA, Jobin Yvon Ltd., Stanmore, Middlesex, UK, 2012.

31. Dawson, W.R. and M.W. Windsor, *Fluorescence yields of aromatic compounds.* The Journal of Physical Chemistry, 1968. **72**(9): p. 3251-3260.